\documentclass[acus]{JAC2000}

\usepackage{graphicx,epsfig,psfig}

\begin{document}
\title{\flushright{M05}\\[15pt] \centering Precision Measurement
of $g_\mu-2$ at BNL}

\author{O.~Rind$^1$, H.N.~Brown$^2$, G.~Bunce$^2$, R.M.~Carey$^1$,
P.~Cushman$^{9}$, G.T.~Danby$^2$,\\ 
P.T.~Debevec$^7$, M.~Deile$^{11}$, H.~Deng$^{11}$, W.~Deninger$^7$, 
S.K.~Dhawan$^{11}$, V.P.~Druzhinin$^3$,\\
L.~Duong$^{9}$, E.~Efstathiadis$^1$, F.J.M.~Farley$^{11}$, 
G.V.~Fedotovich$^3$, S.~Giron$^{9}$,\\
F.~Gray$^7$, D.~Grigoriev$^3$, M.~Grosse-Perdekamp$^{11}$,
A.~Grossmann$^6$, M.F.~Hare$^1$,\\
D.W.~Hertzog$^7$, V.W.~Hughes$^{11}$, M.~Iwasaki$^{10}$,
K.~Jungmann$^6$, D.~Kawall$^{11}$,\\
M.~Kawamura$^{10}$, B.I.~Khazin$^3$, J.~Kindem$^{9}$, F.~Krienen$^1$,
I.~Kronkvist$^{9}$, R.~Larsen$^2$,\\ 
Y.Y.~Lee$^2$, I.~Logashenko$^{1,3}$, R.~McNabb$^{9}$, W.~Meng$^2$, 
J.~Mi$^2$, J.P.~Miller$^1$, W.M.~Morse$^2$,\\
D.~Nikas$^2$, C.J.G.~Onderwater$^7$, Y.~Orlov$^4$, C.S.~\"{O}zben$^2$,
J.M.~Paley$^1$, C.~Polly$^7$,\\
J.~Pretz$^{11}$, R.~Prigl$^2$, G.~zu~Putlitz$^6$, S.I.~Redin$^{11}$,
B.L.~Roberts$^1$, N.~Ryskulov$^3$,\\ 
S.~Sedykh$^7$, Y.K.~Semertzidis$^2$, Yu.M.~Shatunov$^3$, 
E.P.~Sichtermann$^{11}$, E.~Solodov$^3$,\\
M.~Sossong$^7$, A.~Steinmetz$^{11}$, L.R.~Sulak$^1$,
C.~Timmermans$^{9}$, A.~Trofimov$^1$,\\ 
D.~Urner$^7$, P.~von~Walter$^6$, D.~Warburton$^2$, D.~Winn$^5$, 
A.~Yamamoto$^8$, D.~Zimmerman$^{9}$\\ 
\vspace{\baselineskip}\\
$^1$Department of Physics, Boston University, Boston, MA 02215, USA\\
$~^2$Brookhaven National Laboratory, Upton, NY 11973, USA\\
$~^3$Budker Institute of Nuclear Physics, Novosibirsk, Russia\\
$~^4$Newman Laboratory, Cornell University, Ithaca, NY 14853, USA\\
$~^5$Fairfield University, Fairfield, CT 06430, USA\\
$~^6$Physikalisches Institut der Universit\"{a}t Heidelberg, 69120 Heidelberg, Germany\\
$~^7$Department of Physics, University of Illinois at Urbana-Champaign, IL 61801, USA\\
$~^8$KEK, High Energy Accelerator Research Organization, Tsukuba, Ibaraki 305-0801, Japan\\
$~^{9}$Department of Physics, University of Minnesota, Minneapolis, MN 55455, USA\\
$~^{10}$Tokyo Institute of Technology, Tokyo, Japan\\
$~^{11}$Department of Physics, Yale University, New Haven, CT 06520, USA
}

\maketitle

\begin{abstract}
The Muon $(g-2)$ Experiment (E821) at Brookhaven National Laboratory
(BNL) has measured the anomalous magnetic moment of the positive muon
to an unprecedented precision of 1.3 parts per million.  The result,
$a_{\mu^+} = \frac{g-2}{2} = 11\ 659\ 202(14)(6) \times 10^{-10}$, is
based on data recorded in 1999 and is in good agreement with previous
measurements.  Upcoming analysis of data recorded in 2000 and 2001
will substantially reduce the uncertainty on this measurement.
Comparison of the new world average experimental value with the most
comprehensive Standard Model calculation, $a_\mu(SM) = 11\ 659\
159.6(6.7) \times 10^{-10}$, yields a difference of $a_\mu({\rm
exp})-a_\mu({\rm SM}) = 43(16) \times 10^{-10}$.
\end{abstract}

\section{Introduction}
Lepton anomalous magnetic moments arise from purely quantum mechanical
effects, predominantly through higher order corrections to the
$ll\gamma$ vertex.  Precision measurements of these quantities have
played an important role in the development of quantum field theory
throughout the last century and continue to test the limits of our
theoretical knowledge even today.  Currently, the electron anomaly is
one of the most precisely measured quantities in physics, known to an
extraordinary accuracy of 4 parts per billion (ppb)\cite{a_e-exp}.
Even at this level, it includes contributions from QED loop
corrections only.  As a result, it currently provides the best
determination of the fine structure constant, under the assumption of
the validity of QED.

The muon anomaly, on the other hand, has now been measured to a level
of 1.3 parts per million (ppm)\cite{e821-99}, as discussed in this
note.  Although this measurement is about 350 times less precise than
that of the electron, it is already far more sensitive to hadronic and
electroweak loop contributions, as well as any new, non-Standard Model
effects.  This is because the strength of such virtual loop terms is
generally proportional to the square of the relevant mass scale, thus
giving an enhancement of $m_\mu^2/m_e^2 \approx 40,000$ in the
contribution to the muon relative to the electron.  In essence, the
higher mass scale of the muon provides a much more effective probe of
short-distance phenomena.

The anomalous magnetic moment is generally written as $a =
\frac{g-2}{2}$, where the $g$-factor relates the magnetic moment of
the particle to its spin, $\vec{\mu} = g(e/2mc) \vec{S}$.  In the SM,
the contributions to the muon anomaly can be written as the sum of
three general classes of diagrams: 
\begin{equation}
a_\mu({\rm SM}) = a_\mu({\rm QED}) + a_\mu({\rm Hadronic}) +
a_\mu({\rm Weak}).
\end{equation}
Table~\ref{tb:SM} gives a breakdown of these terms with their relative
contributions both to the value and uncertainty of $a_\mu$(SM).  As in
the electron case, the QED term dominates; however, hadronic and even
weak contributions already come into play at the ppm level.  Indeed,
the previous measurement of $a_\mu$ conducted at CERN in the 1970's
had already demonstrated the presence of the hadronic contribution
with an uncertainty of 7.3 ppm\cite{cern3}, mostly statistical.  One
of the initial design goals of the BNL experiment was a factor of 20
improvement (0.35 ppm) on the CERN result, giving a more than
$3\sigma$ sensitivity to the weak contribution.  It is interesting to
note that this measurement has now been conducted four times (three
times at CERN) and each time the result has been sensitive to
theoretical contributions at a new level of computation.

\begin{table}
\caption{Standard Model contributions to $a_\mu$\cite{czar-mar99,dh98}.}
\begin{tabular}{lrr}
Term               & Value $(\times 10^{-11})$ & Rel. Cont. (ppm) \\
\hline
$a_\mu$(QED) 	   & 116\ 584\ 705.7(2.9) & $10^6 \pm 0.02$ \\
$a_\mu$(Hadronic)  & 6739(67) &  $57.79 \pm 0.57$ \\
$a_\mu$(Weak)      & 151(4) & $1.30 \pm 0.03$ \\ \hline
$a_\mu$(SM)        & 116\ 591\ 596(67) & $\pm 0.57$
\end{tabular} 
\label{tb:SM}
\end{table}

The uncertainty in the theoretical value of $a_\mu$ is currently
dominated by knowledge of the hadronic term.  Because of the
non-perturbative aspects of low energy QCD, evaluation of this term is
not possible from first principles and requires input from experiment,
specifically $e^+e^-\rightarrow$ {\it hadrons} (and, recently,
hadronic $\tau$-decay) cross-sections down to the pion production
threshold.  Measurement of these cross-sections is absolutely crucial
to the interpretation of any $a_\mu$ result.  Accordingly, improvement
is expected soon from the experimental programs in
Novosibirsk\cite{cmd2} and Beijing\cite{bepc}.  Their work is
discussed elsewhere in these conference proceedings, along with goals
in the longer term\cite{pepn-lowe}.

A recent review of the current state of the theory can be found in
reference \cite{czar-mar01} and the citations therein, as well as
other presentations in this conference session\cite{pepn-theory}.

\section{The BNL Experiment}

The experimental principle in the Brookhaven experiment is similar in
concept to that of the final CERN experiment\cite{cern3}.  A polarized
muon beam is stored in a highly uniform, circular, dipole magnet and
the decay rate of muons in flight is measured with high precision.  In
the presence of a magnetic and electric field, the muon spins precess
in the lab frame with the angular frequency
\begin{equation}
\label{eq:omegaa}
\vec{\omega}_a = -\frac{e}{m_\mu}\left[ a_\mu \vec{B} - \left(
 a_\mu - \frac{1}{\gamma^2-1}\right)\vec{\beta}\times\vec{E}\right].
\end{equation}
In this expression, $\vec{\omega}_a$ is the angular frequency of the
spin vector relative to the momentum vector.  This frequency is
proportional to $a_\mu$ itself, not $g$, enabling a higher precision
direct measurement of the anomaly.  The beam is focused in the ring
vertically using an electrostatic quadrupole field.  At a specific
``magic'' momentum, the second term in equation~\ref{eq:omegaa} drops
out and the spin precession is unaffected by the focusing electric
field.  This momentum, $p_\mu = 3.094$ GeV/c $(\gamma = 29.3)$, sets
the scale of the experiment, some parameters of which are shown in
table~\ref{tb:expar}.

Measurement of $a_\mu$ thus requires simultaneous determinations of
both $\omega_a$ and the magnetic field.  In practice, the magnetic
field is determined using an NMR system that measures the free proton
precession frequency in the same magnetic field seen by the muons.
The anomaly is then extracted through the relation
\begin{equation}
\label{eq:amu}
a_\mu = \frac{\omega_a/\omega_p}{\mu_\mu/\mu_p - \omega_a/\omega_p}
\end{equation}
where the only external input is the ratio of muon to proton magnetic
moments, $\mu_\mu/\mu_p = 3.18334539(10)$\cite{lambda}.  This
expression enables a natural separation of the measurement into two
independent analyses, one of the field and one of the muon spin
precession.  Any experimenter bias can be eliminated by maintaining
secret offsets between the two analysis groups.  Once the analyses are
complete, the results are frozen, the offsets are revealed and only
then is the value of $a_\mu$ determined.

\begin{table}
\caption{E821 parameters}
\begin{tabular}{ll}
Parameter & Value \\ \hline
$B_0$ & 1.45 T \\
Orbit Radius & 7.112 m \\
Storage Region Diameter & 9 cm \\
Momentum & 3.09 GeV/c \\
$\gamma$ & 29.3 \\
$\gamma\tau$ & 64.4 $\mu s$ \\
Cyclotron Frequency & 6.70 MHz \\
g-2 Frequency & 0.229 MHz \\
Field Index (n) & 0.137 \\
Horizontal Tune $(\nu_h)$ & 0.93 \\
Vertical Tune $(\nu_v)$ & 0.37 \\
AGS storage & $6 \times 10^{13}$ protons \\
AGS rep rate & 0.38 Hz \\
Beam width $(\sigma)$ & 25 ns
\end{tabular}
\label{tb:expar}
\end{table}

\subsection{The muon beam}

The Alternating Gradient Synchrotron (AGS) delivers up to $6 \times
10^{13}$ protons, at an energy of 24 GeV, in 12 bunches (6 in 1999)
every 2.5 seconds.  The bunches are extracted every 33 ms and directed
onto a nickel target.  Pions of $\approx 3.1$ GeV/c are transported
from the target down a 116 m beamline where about 50\% decay into
muons.  Forward-going muons are then momentum selected for injection
into the muon storage ring with a polarization of $\approx 96\%$.

The muons enter the ring through a hole in the magnet yoke and pass
through a field-free region supplied by a DC superconducting inflector
magnet\cite{inflector}.  The fringe field of the inflector magnet is
contained by a superconducting shield designed to limit its effect to
$\sim 1$ ppm at 2 cm.  Upon exiting the inflector channel, the muons
are in an orbit offset by 7.7 cm from the center of the storage
region.  A fast, pulsed magnet provides the $\approx 11$ mrad kick
needed to move the beam onto a central orbit.  This kicker magnet
reduces the storage ring field locally by 0.016 T for $\approx 450$ ns
with less than a 0.1 ppm residual effect after 20 $\mu$s.

This direct muon injection technique is one of the major technical
improvements of this experiment compared to the pion injection
technique of the CERN experiment\cite{cern3}, allowing for more
efficient injection with greatly reduced background.  Approximately
one muon is stored for every $10^9$ protons on target.

\subsection{The storage ring magnet}

\begin{figure*}[!htb]
\centering
\includegraphics*[width=150mm]{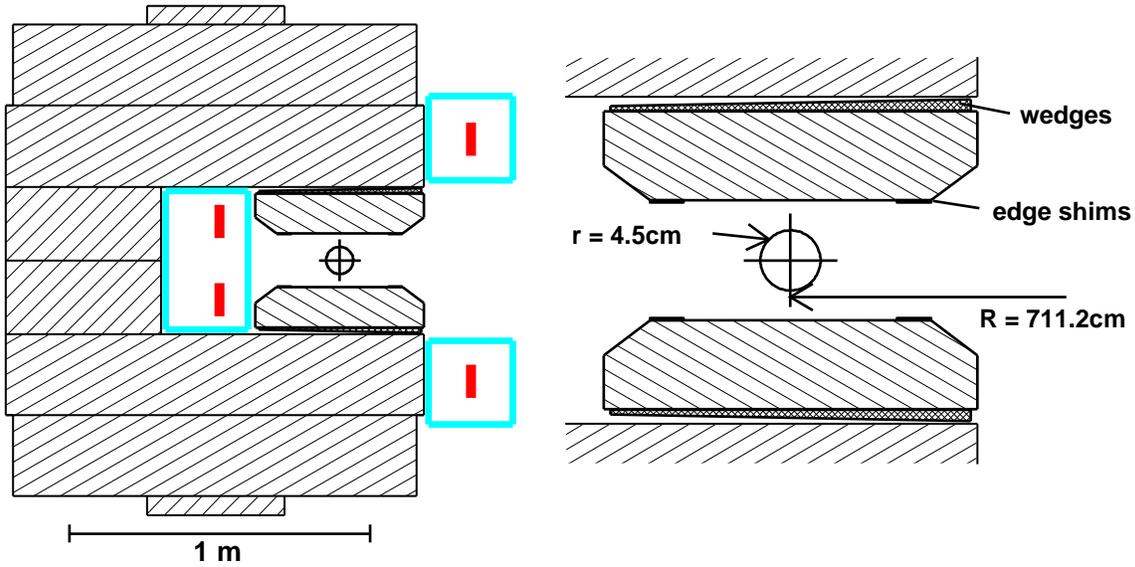}
\caption{Cross-section of the $(g-2)$ storage ring magnet with an
expanded view of the magnet gap region.  The superconducting coils run
perpendicular to the page in three cryostat boxes as shown.  The 9 cm
diameter circular beam storage region is indicated.}
\label{fig:g2magnet}
\end{figure*}

A cross-sectional view of the storage ring magnet\cite{magpaper} is
shown in figure~\ref{fig:g2magnet}.  It is a continuous, superferric,
C-shaped magnet with a radius of 7.112 m at the center of the storage
region.  The field is excited by superconducting coils carrying a
current of 5.2 kA and is shaped by high-precision iron pole pieces.
The pole pieces are 10 degrees long and separated by 75 $\mu$m
Kapton-insulated gaps in order to avoid irregular eddy current
effects.  Vertical air gaps decouple the pole pieces from the yoke
steel and allow the insertion of iron wedges which are used both to
compensate for the natural quadrupole term due to the C-shaped
geometry and to reduce the azimuthal field inhomogeneity.  A series of
edge shims are used to reduce the local field variations over the beam
cross-section and current sheets glued to the pole faces reduce the
variations in the integral field.  Field changes due to ambient
temperature fluctuations are reduced by insulating both the yoke and
pole pieces.  A feedback loop from the NMR system to the magnet power
supply compensates for drifts in the overall dipole term.  Monitoring
and analysis of the magnetic field is discussed in the next section.

The electrostatic quadrupoles, which provide the vertical beam
focusing, are mounted inside the beam vacuum chamber in four
symmetrically placed locations.  Each quadrupole consists of four
plates traversing $39^\circ$ in azimuth and was operated at 24 kV for
up to 1.4 ms.  This provided a weak-focusing field index of $n \simeq
0.137$, sufficiently removed from beam and spin resonances.

The storage aperture is defined by a set of 9 cm diameter circular
collimators.  This circular cross-section reduces the coupling of
higher order field multipoles to the beam distribution.  The
collimators are also used to scrape off the tails of the beam
distribution during the first 15 $\mu s$ after injection, thus
reducing beam losses during the measurement period.  This is
accomplished by lowering the voltage on the inner and bottom
quadrupole plates to shift the central beam orbit by up to several
millimeters.

\section{Magnetic Field Analysis}

The first part of the $a_\mu$ analysis requires a detailed measurement
of the magnetic field averaged over the ensemble of stored muons.  One
of the major advances of the BNL experiment is the ability to map out
the magnetic field {\it in vacuo} throughout the storage region.  This
is done using a hermetically sealed trolley containing a matrix of 17
NMR probes.  The trolley moves on fixed rails inside the vacuum
chamber, measuring about 6000 points in azimuth, every 7 mm.
Uncertainty in the azimuthal position of the trolley contributes a 0.1
ppm systematic error to the field measurement.

\begin{figure}[!htb]
\epsfysize=2in
\centerline{\epsfbox{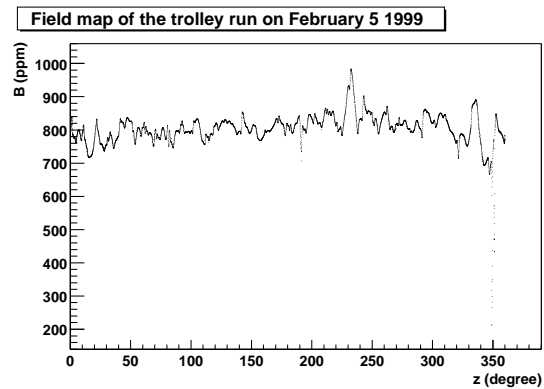}}
\caption{Magnetic field measured with the central trolley probe
relative to an arbitrary reference $B_0$ vs. azimuthal angle around
the storage ring.  The dip due to the inflector occurs near
$350^\circ$.}
\label{fig:fieldmap}
\end{figure}

\begin{figure}[!htb]
\epsfysize=2in
\centerline{\epsfbox{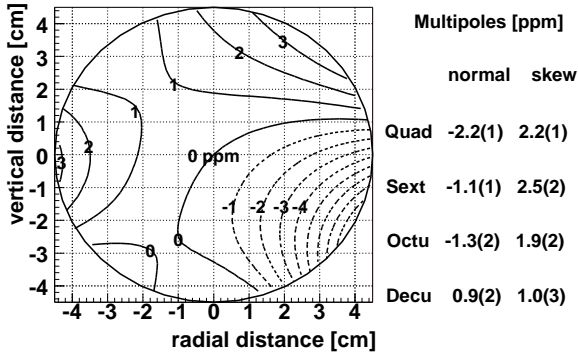}}
\caption{Typical multipole expansion of the field as measured by the
NMR trolley.  The field is averaged over azimuth and one ppm contours
are shown with respect to a central field value of $B_0 = 1.451\ 266$
T.  The circle indicates the beam storage region and the multipole
amplitudes are measured at the 4.5 cm radius.}
\label{fig:bcross}
\end{figure}

Figure~\ref{fig:fieldmap} shows the field mapped by the central
trolley probe around the storage ring.  In 1999, a residual fringe
field in the inflector region caused a dip in the central field,
visible near $350^\circ$, which contributed 0.2 ppm to the field
systematic error.  This effect is also visible in the lower right
corner of figure~\ref{fig:bcross}, which shows a typical field profile
across the storage region, averaged over azimuth.  The inflector was
replaced before the 2000 run, thus eliminating this effect.

Field mappings are conducted every 3 days on average.  In the interim
period, the field is tracked using about 150 NMR probes located in the
upper and lower walls of the vacuum chamber.  The tracking uncertainty
is 0.15 ppm, as determined by comparison of the average field measured
by the fixed probes to that measured by the trolley during each field
mapping.  Before and after data-taking periods, the trolley probes are
calibrated in air against a standard spherical water probe with an
accuracy of 0.2 ppm.  Two largely independent field analyses were
conducted using different selections of NMR probes.  The results
agreed to within 0.03 ppm.

The field integral encountered by the muon beam is studied by tracking
4000 muons for 100 turns through a measured field map.  The
simulation shows that the average field integral over the muon paths
is equivalent, within 0.05 ppm, to the azimuthally averaged field
measurement taken at the beam center.  The radial center of the beam
is known to be $3.7 \pm 1$ mm outside of the central orbit based on
studies of the bunched beam rotation frequency\cite{e821-97}.  The
vertical center is measured to be $2 \pm 2$ mm above the central orbit
using scintillating fiber beam monitors, front scintillator detectors,
and the traceback chamber\cite{design-report}.  These measurements
contribute an additional 0.12 ppm to the uncertainty in $\widetilde
\omega_p$, as shown in table~\ref{tb:B_error}.  The final value is
$\widetilde\omega_p = 61\ 791\ 256 \pm 25$ Hz (0.4 ppm).

\begin{table}
\caption {Systematic errors for the $\widetilde{\omega}_p$ analysis}
\begin{tabular}{lc} 
Source of errors & Size [ppm] \\
\hline
Calibration of trolley probes &  $0.20$ \\
Inflector fringe field & $ 0.20$ \\
Interpolation with fixed probes  & $0.15$ \\
Others $\dagger$ & $0.15$\\
Uncertainty from muon distribution & $0.12$\\
Trolley measurements of $B_0$ & $0.10$ \\
Absolute calibration of standard probe~~~ & $0.05$\\
\hline
Total systematic error on $\widetilde{\omega}_p$ & $0.4\,~$ \\
\end{tabular}
{\small $\dagger$ higher multipoles, trolley temperature and its power supply 
voltage response, and eddy currents from the kicker.}
\label{tb:B_error}
\end{table}

\section{Spin Precession Analysis}

The spin precession frequency is obtained from the muon decay time
spectrum.  In the muon rest frame, the parity violating nature of the
weak decay $\mu^+ \rightarrow e^+ \nu_e \bar{\nu}_\mu$ causes the
positrons to be emitted preferentially along the muon spin direction.
When boosted into the lab frame, this results in a strong correlation
between the positron energy and the angle between the muon spin and
momentum vectors.  The decay positrons, ranging in energy from 0-3.1
GeV, spiral in towards the center of the ring where they are detected
by 24 lead/scintillating fiber calorimeters\cite{calopaper} placed
symmetrically along the inner wall of the vacuum chamber.  The number
of positrons observed above an energy $E_t$ is modulated by the muon
spin precession, yielding a count rate of
\begin{equation}
\label{eq:5par}
N(t) = N_0(E_t) e^{-t/\gamma\tau}[1+A(E_t)\cos(\omega_a t +
\phi_a(E_t))]
\end{equation}
where $\gamma\tau\approx 64.4\ \mu s$ is the dilated muon lifetime.
The normalization, phase and asymmetry all depend on the energy
threshold.  In fitting to this function, the statistical uncertainty
in $\omega_a$ goes as $1/(\sqrt{N}A)$.  In the BNL experiment, a
special scalloped vacuum chamber design ensured that positrons
entering the face of the calorimeter would traverse similar paths
through the chamber wall.  This improves the energy resolution and,
therefore, the asymmetry.  For an energy threshold of 2 GeV, the
asymmetry is $\approx 0.4$.

The calorimeter pulses are sampled by custom-built 400 MHz waveform
digitizers (WFD) which are clocked by the same LORAN-C frequency
receiver used in the NMR system, thus avoiding possible systematics
due to slewing time standards.  Pulses above a predetermined hardware
energy threshold of $\sim 900$ MeV trigger the WFD to record at least
16 8-bit ADC samples (40 ns) on both the fast-rising edge and slower
tail of the pulse.  Single pulses have a typical width of $\sim 5$ ns
and multiple pulses can be resolved if their separation exceeds 3 to 5
ns.  Pulses with energies below the hardware threshold can therefore
be seen if they appear within the sampling time around a trigger
pulse.  This property of the WFD is useful for pileup studies as
described below.

\begin{figure}[!htb]
\center
\epsfig{file=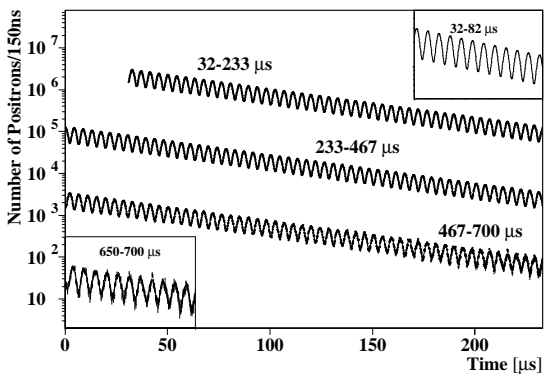, width=3.5in}
\caption{Observed decay positron spectrum in 1999.}
\label{fig:data99}
\end{figure}

The decay spectrum from the 1999 run, containing $\sim 1$ billion
measured positrons, is shown in figure~\ref{fig:data99}.  With this
large a data sample, several effects that cause a deviation from
the ideal functional form of equation~\ref{eq:5par} become
statistically significant.  Determination of an appropriate functional
form is, therefore, an important experimental challenge.  Four
independent analyses were conducted with somewhat different
approaches.  All were forced to confront several common issues which
are enumerated below.

\begin{enumerate}

\item{{\bf Pileup Effects}: 
With the higher data rate in 1999, positron pileup in the calorimeters
became a relevant issue in the analysis.  Pileup events occur when two
positrons arrive within the 3-5 ns deadtime interval of the pulse
finding algorithm.  This changes the number of counts above threshold
in a rate-dependent fashion $(\propto (dN/dt)^2)$.  Thus, the effect
is largest early in the fill and dies out exponentially with half the
dilated muon lifetime.  The effect on the count rate can be positive
or negative: two pulses {\em below} threshold can overlap to mimic a
single pulse above threshold, thus adding a count, or two pulses {\em
above} threshold can overlap to mimic a single pulse, thus losing a
count.

Aside from affecting the count rate, the pileup pulses arrive with a
different phase than $\phi_a$ in equation~\ref{eq:5par}.  Since the
phase is highly correlated with $\omega_a$ in the fit, failure to take
the pileup into account can lead to a shift in the measured frequency.
The pileup functional form can be added to the fit, but the strong
correlation with $\phi_a$ requires that the pileup phase be fixed.
The pileup phase, however, is difficult to measure.

Another approach is to correct the spectrum for pileup effects prior
to fitting.  This can be done by subtracting a pileup spectrum that is
statistically constructed from the data itself.  The technique is
based on the presumption that the likelihood of a second pulse
arriving within the deadtime window around the first pulse is equal to
the likelihood that it will arrive within a similar time interval a
few nanoseconds earlier or later.  It is made possible due to the
extended pulse sampling provided by the WFD, as described above.

Two equivalent software methods are used to correct the time spectrum.
One measures the effect of artificially increasing the deadtime and
uses the result to extrapolate back to the zero deadtime case.  The
other constructs a pileup spectrum out of pulses appearing within a
fixed time window on the tail of each trigger pulse, then subtracts it
from the original time spectrum.  Both methods can only fully correct
data sets with energy thresholds of at least twice the hardware
trigger threshold.  This lead to a choice of $E \ge 2$ GeV in the
analysis.  With this energy selection, the pileup level is about 1\%
at the beginning of the fits.

Pileup pulses below detectable energy thresholds are not corrected by
this procedure.  Since these pulses affect both the baselines and the
pulse heights, they do not change the pulse energies on average.  They
can, however, change both the measured phase and asymmetry.  The
asymmetry is more sensitive to this effect, so it is used to set a
limit on the shift in $\omega_a$.}

\item{{\bf AGS background}: 
Imperfect proton extraction from the AGS sometimes leads to particles
coming down the beamline and entering the storage ring during the
$\sim 1$ ms data collection period.  Some of these particles, mostly
positrons, create background pulses in the detectors which can enter
the data sample.  These pulses appear with a specific time structure,
defined by the 2.694 $\mu s$ AGS cyclotron period, and a specific
azimuthal distribution around the ring, which can be exploited to
enhance their effect and measure the level of contamination.  In 1999,
the relative AGS background level was $\sim 10^{-4}$ which,
simulations show, leads to an uncertainty of $\pm 0.1$ ppm in
$\omega_a$.  In subsequent runs, this background level has been
reduced by employing a fast sweeper magnet to close off the beamline,
downstream of the target, once the main bunch has passed.  Monitoring
of the background level has also improved by periodically suppressing
the quadrupole voltages for a fill to look for background without the
presence of stored beam.}

\item{{\bf Muon Losses}:
Muon beam losses during the data collection period can distort the
exponential decay form.  Such losses are minimized by controlled
scraping of the beam before the start of the fit, as described
above.  Remaining losses are taken into account by multiplying the
functional form of equation~\ref{eq:5par} with an extra loss term:
\begin{equation}
\label{eq:muloss}
l(t) = 1 + n_l e^{-t/\tau_l}
\end{equation}
These losses are also studied using coincident signals in the
front scintillation counters mounted on groups of three adjacent
calorimeters.}

\item{{\bf Gain and Timing Shifts}:
Detector gain and timing stability are monitored with a pulsed laser
system to strict tolerances.  Drifts in gain are also observable
through changes in the positron energy spectra.  Timing shifts are
stable to within 20 ps over the first 200 $\mu$s of the fit (0.1~ppm)
while gain changes are below 0.1\% for all but two of the detectors.
Two of the analyses apply a gain correction, one through the use of a
time-dependent energy threshold and one by incorporating the gain
dependence into the fitting function.}

\item{{\bf Coherent Betatron Oscillations}: 
The inflector aperture is smaller than the storage ring aperture, so
the phase space for betatron oscillations, defined by the acceptance
of the storage ring, is not filled.  With ideal injection, this leads
to a modulation of the horizontal and vertical beam widths at a
characteristic frequency defined by the field index.  However, because
the muon kicker was forced to operate slightly below its design value,
the horizontal injection kick was insufficient to place the beam onto
the ideal orbit.  This resulted in oscillation of the beam centroid
around the central orbit at the betatron frequency.  These coherent
betatron oscillations (CBO) are observed directly using a set of
scintillating fiber beam monitors as shown in figure~\ref{fig:cbo}.
Note that the oscillation frequency is determined by the beam tune ($f
\approx f_c(1-\sqrt{1-n})$, with $f_c$ the cyclotron frequency) and
can be changed using different quadrupole settings.  In the recently
completed 2001 run, two different field indices were used in order to
study this effect further.

The beam oscillation is also visible in the positron time spectrum
because the detector acceptance is a function of the muon decay
position.  Fourier analysis of the positron data yields a frequency of
$\omega_b/2\pi = (470.2 \pm 0.2\ {\rm kHz})$.  This effect dies out
slowly, with a time constant of $\sim 100\ \mu s$ and can be
effectively taken into account by modulating the fit function with a
Gaussian envelope:
\begin{equation}
\label{eq:cbo}
b(t) = 1 + A_b e^{-t^2/\tau_b^2}\cos(\omega_b t + \phi_b)
\end{equation}
The phase of the CBO changes by $2\pi$ going around the ring, so its
effect is strongly reduced when all the detector spectra are summed
together before fitting.

\begin{figure}[!htb]
\epsfysize=2.8in
\centerline{\epsfbox{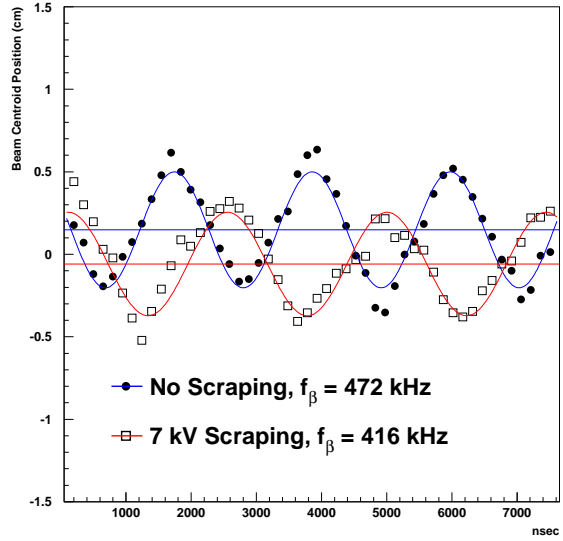}}
\caption{Turn-by-turn evolution of the beam centroid, shortly after
injection, as measured by a scintillating fiber beam monitor at a
fixed azimuthal position in the storage ring.  During the beam
scraping period (see text), the quadrupole plates operate with
asymmetric voltages, thus changing the betatron tune.}
\label{fig:cbo}
\end{figure}
}

\item{{\bf Bunched Beam}:
The beam enters the storage ring with a 25 ns bunch width and
debunches over time due to the $\sim 0.6\%$ momentum spread.  The
measured decay rate is strongly modulated by this bunching effect in
the early part of the fit, but can be eliminated by uniformly
randomizing the start time for each fill over the range of one
cyclotron period.}

\end{enumerate}

The internal consistency of the four different analyses was verified
through a variety of statistical tests.  The results agreed to within
0.3 ppm, which is within the statistical variation expected from the
use of slightly different data sets.  The final value is a weighted
sum of the four results with an error accounting for the strong
correlations due to data overlap: $\omega_a/2\pi = 229072.8 \pm 0.3$
Hz (1.3 ppm).  This number includes a correction of $+0.81 \pm 0.08$
ppm due to (a) the residual effects of the $\vec{\beta}\times\vec{E}$
term in equation~\ref{eq:omegaa} for beam particles off the magic
momentum and (b) the effect of vertical betatron oscillations tilting
the instantaneous angle between the spin and momentum vectors.  The
systematic errors resulting from all the issues discussed above are
summarized in table~\ref{tb:omegaa_error}.  The overall error is still
dominated by statistics.

\begin{table}
\caption {Systematic errors for the $\omega_a$ analysis.}
\begin{tabular}{lc} 
Source of errors & Size [ppm] \\
\hline
Pileup & $0.13$\\
AGS background  & $0.10$\\
Lost muons & $0.10$\\
Timing shifts & $0.10$\\
E field and vertical betatron oscillation ~~~~&  $0.08$ \\
Binning and fitting procedure & $0.07$\\
Coherent betatron oscillation & $0.05$\\
Beam debunching/randomization & $0.04$\\
Gain changes & $0.02$\\
\hline
Total systematic error on $\omega_a$ & $0.3$\\
\end{tabular}
\label{tb:omegaa_error}
\end{table}

\section{Results and Outlook}

\begin{figure}[!htb]
\epsfysize=2in
\centerline{\epsfbox{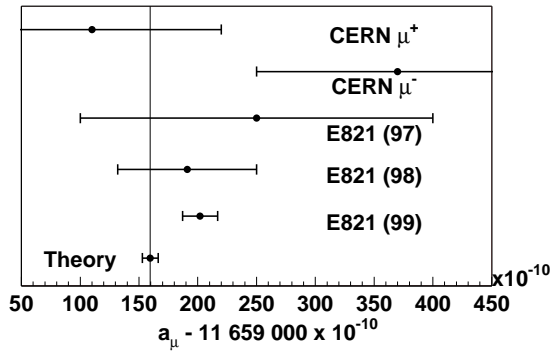}}
\caption{The five most recent measurements of $a_\mu$ and the Standard
Model prediction from a recent, comprehensive calculation~\cite{dh98}.}
\label{fig:history}
\end{figure}

Once the $\omega_p$ and $\omega_a$ analyses were finalized, separately
and independently, the value of $a_\mu$ was calculated using
equation~\ref{eq:amu}.  The result is $a_{\mu^+} = 11\ 659\ 202(14)(6)
\times 10^{-10}$.  This agrees with previous measurements, as shown in
figure~\ref{fig:history}.  The difference between the weighted mean of
the experimental results, $a_\mu({\rm exp}) = 11\ 659\ 203(15) \times
10^{-10}$ (1.3 ppm), and the Standard Model value from
table~\ref{tb:SM} is
\begin{equation}
a_\mu({\rm exp}) - a_\mu(SM) = 43(16) \times 10^{-10}
\end{equation}
where the experimental and theoretical uncertainties were added in
quadrature. 

Many have speculated upon the possible significance of this deviation
from the theoretically expected value.  The experimental result is
certainly expected to improve in the future.  The data set from the
run conducted in early 2000 has approximately four times the
statistics of the 1999 data set.  In 2001, the experiment reversed
polarity and ran with negative muons, collecting a data sample with
about three times the 1999 statistics.  Analysis of these data sets is
now underway and should carry the experiment a long way towards its
stated goal of 0.35 ppm error on the muon anomalous magnetic moment.
In this endeavor, the contribution from measurements at low energy
$e^+e^-$ collider facilities to the theoretical interpretation of the
result cannot be overstated.  We eagerly await the new measurements
which are now on the horizon.

\section{Acknowledgements}

It is a pleasure to thank the organizers of the {\em $e^+e^-$ Physics
at Intermediate Energies Workshop} for their invitation and kind
hospitality.  This work was supported in part by the U.S.  Department
of Energy, the U.S.  National Science Foundation, the German
Bundesminister f\"{u}r Bildung und Forschung, the Russian Ministry of
Science, and the US-Japan Agreement in High Energy Physics.

\end{document}